\pdfoutput=1

\documentclass[prl,reprint,aps,superscriptaddress,nobibnotes,nofootinbib]{revtex4-2}

\usepackage[utf8]{inputenc}
\usepackage{mathptmx}
\usepackage{amsmath}
\usepackage{amssymb}
\usepackage[tight]{units}
\usepackage{color,graphicx}
\usepackage[colorlinks,citecolor=blue,urlcolor=blue,linkcolor=blue]{hyperref}
\usepackage[normalem]{ulem}
\usepackage{enumitem}
\usepackage{mathtools}
\usepackage{braket}

\usepackage[dvipsnames]{xcolor}

\begin{document}
\title{Testing spin-dependent dark matter interactions with lithium aluminate targets in CRESST-III}
\newcommand{\mpi}{\affiliation{Max-Planck-Institut f\"ur Physik, 80805 M\"unchen, Germany}}
\newcommand{\coimbra}{\affiliation{Also at: LIBPhys, Departamento de Fisica, Universidade de Coimbra, P3004 516 Coimbra, Portugal}}
\newcommand{\hephy}{\affiliation{Institut f\"ur Hochenergiephysik der \"Osterreichischen Akademie der Wissenschaften, 1050 Wien, Austria}}
\newcommand{\ati}{\affiliation{Atominstitut, Technische Universit\"at Wien, 1020 Wien, Austria}}
\newcommand{\tum}{\affiliation{Physik-Department, Technische Universit\"at M\"unchen, 85747 Garching, Germany}}
\newcommand{\tuebingen}{\affiliation{Eberhard-Karls-Universit\"at T\"ubingen, 72076 T\"ubingen, Germany}} 
\newcommand{\bratislava}{\affiliation{Comenius University, Faculty of Mathematics, Physics and Informatics, 84248 Bratislava, Slovakia}}

\newcommand{\oxford}{\affiliation{Department of Physics, University of Oxford, Oxford OX1 3RH, United Kingdom}}
\newcommand{\wmi}{\affiliation{Also at: Walther-Mei\ss ner-Institut f\"ur Tieftemperaturforschung, 85748 Garching, Germany}}
\newcommand{\lngs}{\affiliation{INFN, Laboratori Nazionali del Gran Sasso, 67010 Assergi, Italy}}
\newcommand{\gssi}{\affiliation{Also at: Gran Sasso Science Institute, 67100, L'Aquila, Italy}}
\newcommand{\cassino}{\affiliation{Also at: Dipartimento di Ingegneria Civile e Meccanica, Università degli Studi di Cassino e del Lazio Meridionale, 03043 Cassino, Italy}}

\mpi
\lngs
\tum
\hephy
\ati
\tuebingen
\oxford

\coimbra
\gssi
\wmi
\cassino

\author{G.~Angloher}
  \mpi

\author{S.~Banik}
  \hephy
  \ati

\author{G.~Benato}
  \lngs 

\author{A.~Bento}
  \mpi
  \coimbra 

\author{A.~Bertolini}
\email[Corresponding author: ]{anbertol@mpp.mpg.de}
  \mpi

\author{R.~Breier}
  \bratislava

\author{C.~Bucci}
  \lngs 

\author{J.~Burkhart}
  \hephy
  \ati

\author{L.~Canonica}
  \mpi 

\author{A.~D'Addabbo}
  \lngs

\author{S.~Di~Lorenzo}
  \lngs

\author{L.~Einfalt}
  \hephy
  \ati
  
\author{A.~Erb}
  \tum
  \wmi
  
\author{F.~v.~Feilitzsch}
  \tum

\author{N.~Ferreiro~Iachellini}
  \mpi  
  
 \author{S.~Fichtinger}
  \hephy
 
\author{D.~Fuchs}
  \mpi  
 
\author{A.~Fuss}
  \hephy
  \ati

\author{A.~Garai}
  \mpi 
  
 \author{V.M.~Ghete}
  \hephy 

\author{S.~Gerster}
  \tuebingen 

\author{P.~Gorla}
  \lngs 

\author{P.V.~Guillaumon}
  \lngs

 \author{S.~Gupta}
 \email[Corresponding author: ]{shubham.gupta@oeaw.ac.at}
  \hephy 

\author{D.~Hauff}
  \mpi 

\author{M.~Ješkovsk\'y}
  \bratislava

\author{J.~Jochum}
  \tuebingen 

\author{M.~Kaznacheeva}
  \tum

\author{A.~Kinast}
  \tum
  
\author{H.~Kluck}
  \hephy
  \ati

\author{H.~Kraus}
  \oxford

\author{A.~Langenk\"amper}
  \tum
  \mpi

\author{M.~Mancuso}
  \mpi
 
 \author{L.~Marini}
  \lngs
  \gssi

\author{L.~Meyer}
  \tuebingen 
  
\author{V.~Mokina}
  \hephy
 
\author{A.~Nilima}
  \mpi 

\author{M.~Olmi}
  \lngs
  
\author{T.~Ortmann}
  \tum

\author{C.~Pagliarone}
  \lngs 
  \cassino

\author{L.~Pattavina}
  \tum
  \lngs

\author{F.~Petricca}
  \mpi 

\author{W.~Potzel}
  \tum 

\author{P.~Povinec}
  \bratislava

\author{F.~Pr\"obst}
  \mpi

\author{F.~Pucci}
  \mpi 
  
\author{F.~Reindl}
  \hephy
  \ati

\author{J.~Rothe}
  \tum
  
\author{K.~Sch\"affner}
  \mpi

\author{J.~Schieck}
  \hephy
  \ati 

\author{D.~Schmiedmayer}
   \hephy
   \ati

\author{S.~Sch\"onert}
  \tum 
  
\author{C.~Schwertner}
  \hephy
  \ati

\author{M.~Stahlberg}
  \mpi

\author{L.~Stodolsky}
  \mpi 

\author{C.~Strandhagen}
  \tuebingen

\author{R.~Strauss}
  \tum

\author{I.~Usherov}
  \tuebingen 

\author{F.~Wagner}
  \email[Corresponding author: ]{felix.wagner@oeaw.ac.at}
  \hephy

\author{M.~Willers}
  \tum 

\author{V.~Zema}
  \mpi

\collaboration{CRESST Collaboration}
\noaffiliation

\begin{abstract}
In the past decades, numerous experiments have emerged to unveil the nature of dark matter, one of the most discussed open questions in modern particle physics. Among them, the CRESST experiment, located at the Laboratori Nazionali del Gran Sasso, operates scintillating crystals as cryogenic phonon detectors. In this work, we present first results from the operation of two detector modules which both have 10.46~g LiAlO$_2$ targets in CRESST-III. The lithium contents in the crystal are $^6$Li, with an odd number of protons and neutrons, and $^7$Li, with an odd number of protons. By considering both isotopes of lithium and $^{27}$Al, we set the currently strongest cross section upper limits on spin-dependent interaction of dark matter with protons and neutrons for the mass region between 0.25 and 1.5~GeV/c$^2$.
\end{abstract}
\maketitle
\section{Introduction}
\label{sec:intro}

The nature of dark matter (DM) is one of the most discussed open questions in modern physics and has been the motivation for numerous experiments in the past decades. DM direct detection experiments aim at measuring the scattering of DM particles directly off a target material to test interaction scenarios of particle DM with standard model (SM) particles 
\cite{PhysRevD.31.3059}. 
A particularly promising DM model are Weakly Interacting Massive Particles (WIMPs) \cite{Jungman_1996}. Direct-detection experiments searching for WIMPs are sensitive to two parameters: the WIMP mass and its effective interaction cross section.
The original WIMP model considers the weak nuclear force as the mediating force between DM and SM. The model is in conflict with the Lee Weinberg bound for light DM 
\cite{PhysRevLett.39.165} but remains valid for other massive mediators and is used as a benchmark model to compare results from different experiments. 

The Cryogenic Rare Event Search with Superconducting Thermometers (CRESST) experiment probes the interaction of DM with scintillating crystals operated as cryogenic calorimeters in a low-background facility at the Laboratori Nazionali del Gran Sasso (LNGS). The experiment is in its third phase (CRESST-III), focusing on sub-GeV/c$^2$ DM searches, using crystals with light nuclei as targets and Transition Edge Sensors (TES) as phonon sensors. With this technology, CRESST-III provides one of the strongest limits for spin-independent interactions with sub-GeV/c$^2$ DM and the strongest under standard assumptions
\cite{PhysRevD.100.102002}.
The cryogenic technology is versatile and allows for changing the target material.
In recent runs multiple materials were employed simultaneously in individual detector modules: calcium tungstate, sapphire, silicon, and lithium aluminate (LiAlO$_2$). The latter was discussed in reference~\cite{abdelhameed_first_2019}: with a very low atomic number and unpaired nuclei, lithium has appealing properties to test light DM with spin-dependent interactions. 
In previous measurements above ground, the CRESST collaboration demonstrated the competitiveness of DM results achieved with lithium targets. LiAlO$_2$ is a suitable target material in particular, because a TES can be deposited directly on the crystal surface, and the CRESST light detectors have a high absorption at the wavelength of its scintillation peak. The motivations behind the detector design were discussed in detail in reference~\cite{abdelhameed_cryogenic_2020}. The inclusion of $^6$Li in the calculation of DM results was studied in reference~\cite{angloher_probing_2022}. 


In the current phase of the experiment CRESST-III is operating two detector modules with LiAlO$_2$ targets. In this paper we present the dark matter search performed on the data acquired between February and August 2021.
We discuss the design choices of the detector module and the experimental setup at LNGS in section~\ref{sec:setup}. The details of the analysis chain are explained in section~\ref{sec:Analysis}. The data sets allow for the calculation of upper limits on the spin-dependent DM-SM cross section. We present these in section~\ref{sec:Results} and conclude the discussion in section~\ref{sec:Conclusion}.


\section{Detector design and experimental setup}
\label{sec:setup}

\begin{figure}[!t]
\centering
\includegraphics[width=\linewidth]{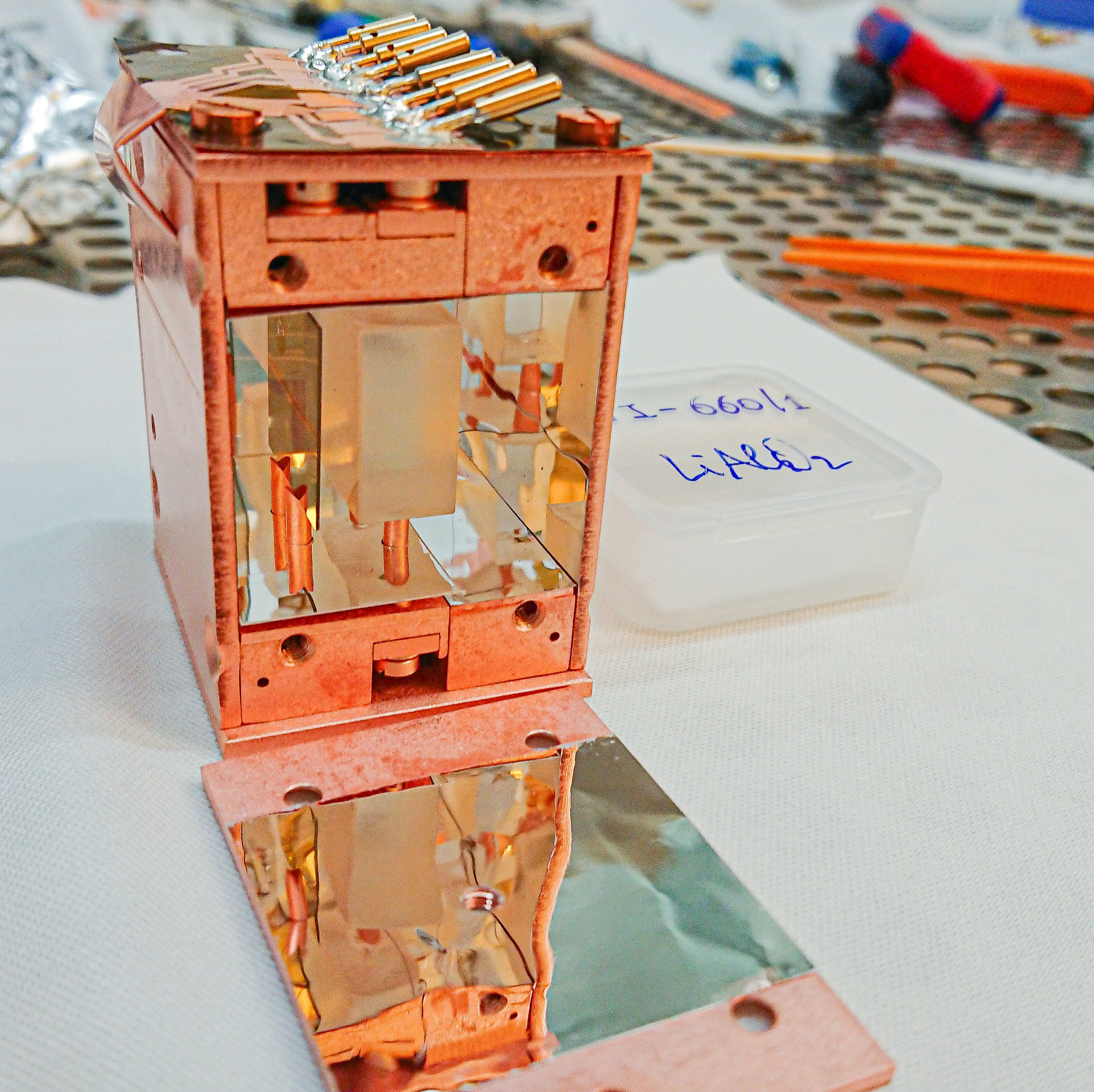} 
\caption{The Li1 detector module. Inside the copper housing, a LiAlO$_2$ crystal (right, transparent) as a target for particle scattering is held by three copper sticks as a target for particle scattering. Next to the crystal an SOS light detector (left, grey) is mounted. The inside of the housing is covered with reflective foil, best visible on the detached side of the module (centre, lower part of the picture).}
\label{fig:module}
\end{figure}

The detector modules "Li1" and "Li2" were identically manufactured in the laboratories of the Max-Planck-Institut f\"ur Physik in Munich. A picture of the Li1 module during the assembly phase is displayed in figure~\ref{fig:module}. 

The modules are constituted by a phonon detector and a light detector, mounted in a copper housing. The phonon detectors feature LiAlO$_2$ absorber crystals, which were provided by the Leibniz-Institut f\"ur Kristallz\"uchtung, and have the dimensions of (2~x~2~x~1)~cm$^3$. The targets have a weight of 10.46~g each. The lithium in LiAlO$_2$ occurs as $^7$Li and $^6$Li with natural abundances of 92.41\% and 7.49\%, respectively \cite{natabundance} while aluminium occurs as $^{27}$Al with a natural abundance of 100\%. LiAlO$_2$ emits scintillation light with an emission maximum at a wavelength of 340~nm \cite{doi:10.7566/JPSJ.86.094201}. A Silicon-On-Sapphire (SOS) substrate of (2~x~2~x~0.04)~cm$^3$ is placed next to the crystal to detect the scintillation light. Both the LiAlO$_2$ crystal and the SOS substrate are equipped with a tungsten TES featuring Al phonon collectors. 

The housing of the detector modules are made from copper with the crystals held in place by three copper sticks. The inner side of the housing is covered with reflective and scintillating foil, a 3M Vikuiti$^\text{TM}$ Enhanced Specular Reflector, to maximise the collection efficiency of scintillation light emitted by the target LiAlO$_2$ crystal. 

The modules were operated next to each other in the CRESST setup at LNGS, which is located below the Gran Sasso massif in central Italy with a rock overburden of 3600~m water equivalent. The vertical muon flux inside the lab was measured to $\sim$~1~counts~/~m$^2$~/~h \cite{PhysRevD.52.3793, PhysRevD.58.092005, Schumann_2019}. Remaining muons are tagged by active muon veto panels around the experiment, which cover the detector location to more than 98\%. Additional shieldings are in place: a polyethylene layer protects the detectors from environmental neutrons. Inside the polyethylene, a lead and a copper layer shield against $\gamma$-rays. Directly surrounding the detector modules, a second polyethylene layer moderates neutrons produced inside the lead and copper. 


For the measurement, the detector modules are cooled with a commercial $^3$He/$^4$He-dilution refrigerator to a base temperature of about 5~mK. The temperature of the TES is stabilised with heating resistors on the holding structure of the detector modules and on the crystal itself, to an operation point within the superconducting phase transition, which is around 15~mK. The heating resistor is also used to periodically induce thermal pules that saturate the TES (control pulses) to measure and stabilise the exact working point within the superconducting transition. Additionally, in between the control pulses, thermal pulses (in the following called “test pulses”) with certain amplitudes (TPA) are sent to monitor the calibration over time \cite{angloher_limits_2005, angloher_commissioning_2009}. The TESs are read out by a SQUID amplifier and continuously digitised with 16 bit precision and 25~kHz sampling frequency.

An $^{55}$Fe source was mounted inside each detector housing to calibrate the detector response to electron recoils. For the calibration of the detector response to nuclear recoils and after the collection of the data set for the calculation of physics results, an AmBe source was put in place, outside the shielding of the experimental setup, to provide a strong neutron flux.

The light detector of Li2 could not be operated, the module could therefore only be read out with one channel, the phonon channel. However, the Li1 module has the scintillation light channel, which enabled the discrimination between electron and nuclear recoils by their individual quenching factor (QF). We use the Li2 module for cross checks of the analysis chain, while the Li1 module provides the performance for competitive DM results.

\section{Data Analysis}
\label{sec:Analysis}



A particle recoil inside the target produces a population of athermal phonons, which spread ballistically over the crystal. They thermalise mostly through scattering with the crystal surface, heating up the crystal. A share of the athermal phonons is collected by the phonon collectors and led to the thermometer. This produces a temperature signal with two components in the TES, corresponding to the athermal and thermal phonons, respectively. The employed TESs are designed such that the athermal component dominates the pulse height and sensitivity of the detector. For small energy depositions, the pulse height scales approximately linearly with the deposited energy. This model was thoroughly described in reference~\cite{probst_model_1995}. For larger recoil energies, saturation effects of the TES cause a flattening of the pulses. In our DM data set we consider only the region of linear pulse height.



The data set of the total measurement is split into a training and a blind data set. 
The blind data set for Li1 accumulates to 2665~h measurement time, for the Li2 module to 2716~h. 
The total exposure of the blind set is 1.161~kg~days (Li1) and 1.184~kg~days (Li2). That of the used training set is 0.153~kg~days (Li1 and Li2 each). 
The analysis, including the event selection, is designed on the training set and applied with no further modification to the blind set. This procedure is recommended within the DM community \cite{baxter_recommended_2021}. 

\begin{figure}[!t]
\centering
\includegraphics[width=\linewidth]{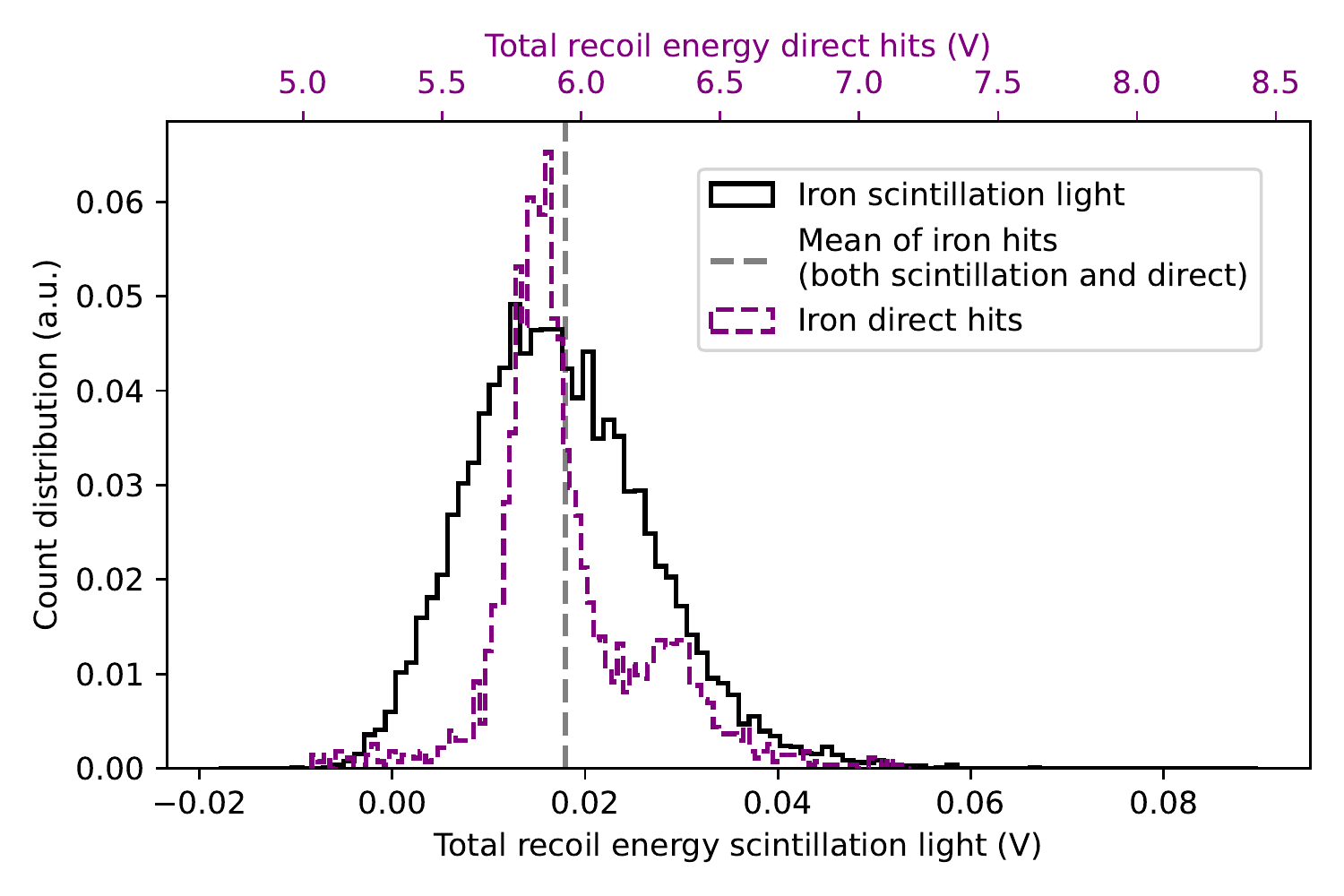}
\caption{Overlay of the normalised Li1 light detector energy count distribution of the iron line from scintillation light (black, bottom x-axis) and the iron line from direct hits (purple dotted, top x-axis). The two x-axes are shifted and scaled, such that the average value of the two iron lines overlap. Their ratio determines the collected light of the target (see text).}
\label{fig:li1_iron}
\end{figure}

\subsection{Trigger and data processing}

As a first step in the detector characterisation we generate a standard event (SEV) for recoil events by averaging a selection of events from a narrow energy interval.
The SEV is then used to create an optimum (matched) filter (OF) that corresponds to the quotient of our SEV and the noise power spectrum in frequency space \cite{gatti_processing_1986, alduino_low_2017}. 
The OF provides the theoretically optimal signal-to-noise ratio for events with the characteristic particle recoil SEV shape, and is applied to the recorded data stream for offline triggering with an optimised trigger threshold. The trigger threshold is calibrated to 1 noise~trigger~/~kg~/~day with the method proposed in reference~\cite{MANCUSO2019492} on the training set data. The triggered events are stored in windows with a length of 16384~samples, where the trigger position is placed at 1/4 of the window size. 
We extract the main shape parameters of each triggered pulse for both the phonon and light channel: e.g. pulse height, onset, rise- and decay time of the values within the pulse window. 
These values are stored for the phonon and light channel individually.
Additionally, we fit each pulse with the SEV plus a third-order polynomial to model the baseline fluctuations, and record the fitted pulse height and root mean square (RMS) deviation of the fit. With a truncated template fit, also weakly saturated pulse heights could be reconstructed.
This is done by scaling the SEV until it properly fits the part of the pulse which is within the linear region of the TES response, reconstructing an amplitude which is higher than the saturated one \cite{Stahlberg:thesis}.

For cross checks and validation purposes, the analysis of the modules was done by independent analysts. For data processing and analysis we used a  collaboration internal package "CAT" 
and the publicly available Python package "Cait" \cite{https://doi.org/10.48550/arxiv.2207.02187}. 

\subsection{Energy calibration}


The energy scale of our detector is calibrated with an iron source ($^{55}$Fe) emitting X-rays. 
The spectral lines for X-ray hits in the target and directly in the light detector are clearly visible in both channels, respectively, and provide recoil-type independent (total) energy scales. 
The scintillation light produced by the iron source shining on the target is visible in the light detector as well, thus a measure of the detected light coming from the crystal and an electron equivalent (ee) energy scale calibration are possible. 

The test pulses are used to fine-tune the slight non-linearities in the transition curve. The TPA values scale similarly to the recoil energy of particle events. The pulse heights of particle events are therefore first translated to equivalent TPA values. Knowing the mean energies 5.89~keV and 6.49~keV corresponding to the K-$\alpha$ and K-$\beta$ shells, respectively, we can convert these values with a linear factor to recoil energies.

The collected scintillation light is the share of energy from an electron recoil in the crystal that is emitted as scintillation light and detected by the light detector. We estimate it for Li1, using the spectral X-ray lines from the iron source. Specifically, we compare the mean amplitude of events registered by the light detector when the iron X-ray is absorbed in the crystal (black line in figure~\ref{fig:li1_iron}) to that of events where the X-ray is directly absorbed by the light detector (purple dotted line in figure~\ref{fig:li1_iron}). With this procedure, we measure the value~(0.302~$\pm$~0.001)\%.

\label{subsec:Li1_analysis}
\begin{figure*}[!t]
\centering
\includegraphics[width=\textwidth]{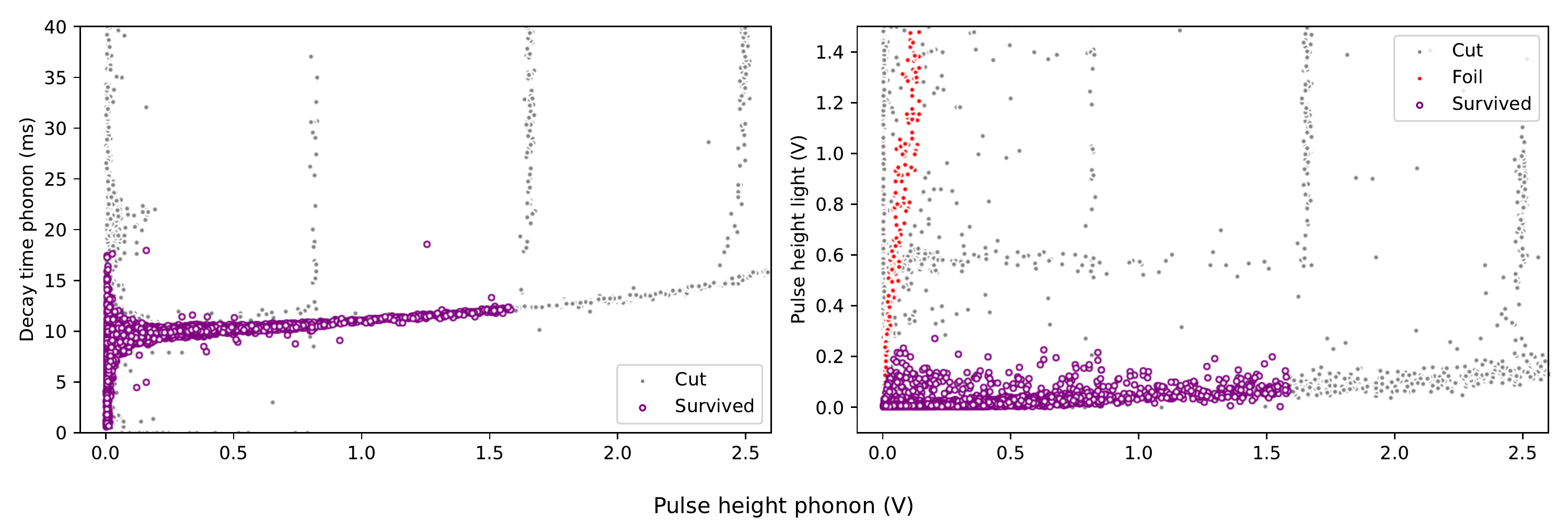}
\caption{Visualisation of the surviving (purple) and cut (grey) events in the Li1 DM data set. (left) The distribution of decay times in the phonon channel over pulse heights in the phonon channel. The band of recoil events is clearly visible and mostly distinct from the artefact events. For pulse heights below 0.2~V the band widens, which degrades the discriminating power of quality cuts. (right) The distribution of pulse heights in the light channel versus the corresponding pulse height in the phonon channel. Again, the band of particle recoils is clearly visible. For low phonon pulse heights the event class of foil events appears: due to their high pulse height in the light channel, higher than for regular target recoils, these events can be rejected as background. In both pictures the vertical event bands, as well as the secondary horizontal event bands, are SQUID resets caused by high energetic recoils.}
\label{fig:li1_mainpars}
\end{figure*}

\subsection{Event selection} \label{sec:eventselection}

We apply several cuts to the events in order to reject non-physical pulses caused e.g.~by earthquakes or human activity inside the laboratory. To develop such cuts, we first select time periods where the detectors were operated in stable conditions:
\begin{itemize}
    \item We exclude periods of time when the detector is out of its operating point. To do so we remove periods where the height of control pulses is not within 3~$\sigma$ of its mean value. 
    \item We calculate the average rate within all ten-minute-intervals of the measurement. We exclude time intervals with a rate not within 3~$\sigma$ of the calculated value.
\end{itemize}
We then use the data from the stable periods to develop quality cuts on the pulse shape parameters which are designed with the goal to keep only events with particle recoils with a correct energy reconstruction.





Finally, we apply an anti-coincidence cut, taking advantage of the muon veto panels, which trigger and record the time stamps of incoming muons.
For each trigger of a muon panel we exclude a window of +10/-5~ms. The muon veto triggers with 4.52~Hz, most of which are dark counts. The muon veto cut removes 6.82$\%$ of the events and 6.79$\%$ of the exposure. The expected percentage of event removed due to random coincidences is (6.79~$\pm$~0.23)$\%$. Similar observations were made for the Li2 module. 

In the same run, CRESST-III operated ten detector modules independently, mounted inside the same holding structure. Due to their low interaction probability, DM recoils are expected to be seen only in single modules (multiplicity 1). Other particle recoils or environment-induced energy depositions can feature a higher multiplicity. Therefore we apply an anti-coincidence cut on the multiplicity of events: for each trigger in another detector module, we exclude a window of +10/-10~ms in the Li1 and Li2 detector. This cut removes 0.93~h runtime in Li1, which is 0.0387$\%$ of the exposure, and two events from the Li1 blind data set. Also for Li2 a negligible share of exposure was removed, and no additional events were rejected by this cut.

\begin{figure}[!t]
\centering
\includegraphics[width=\linewidth]{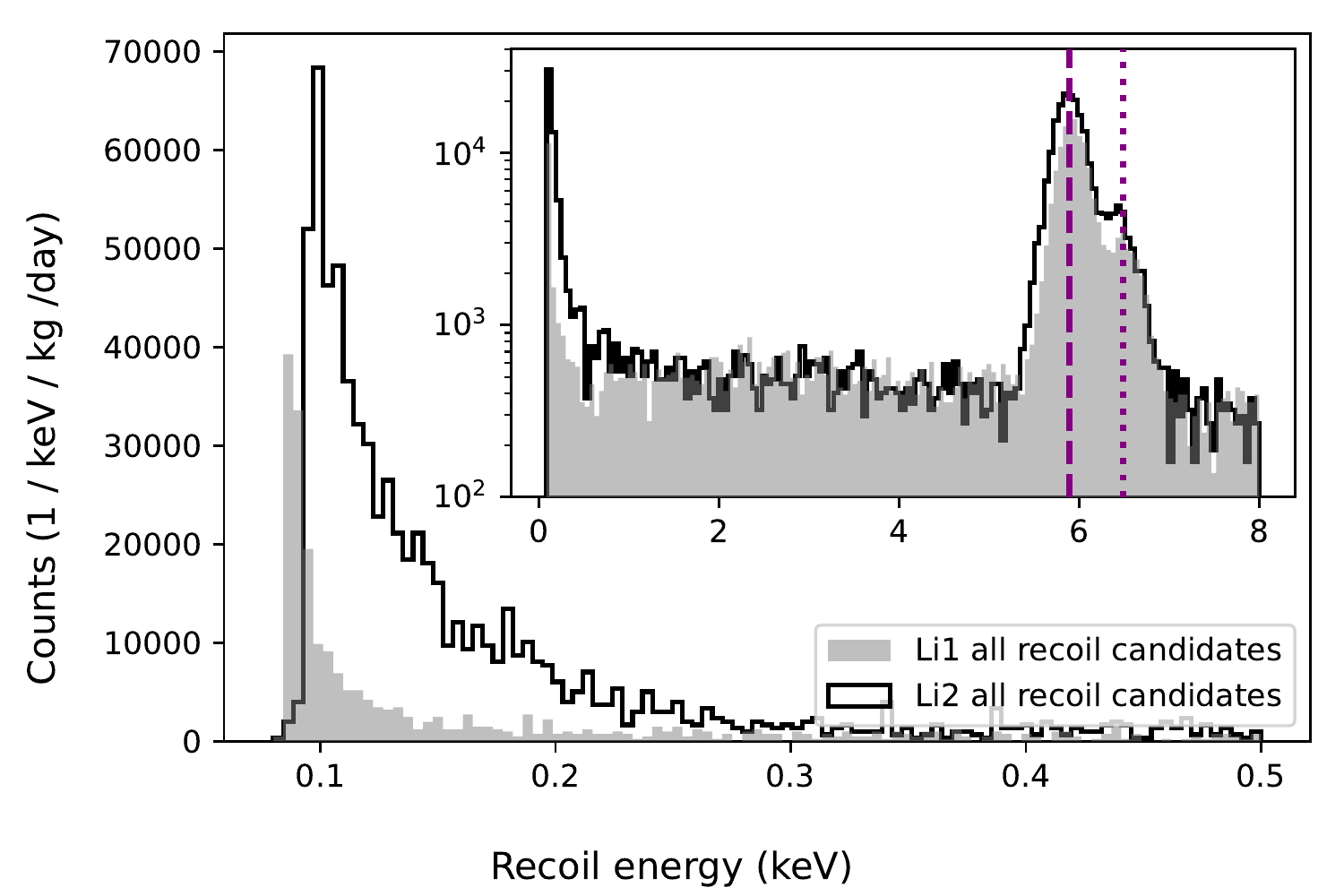}
\caption{Recoil energy spectrum for the Li1 (grey) and Li2 (black) modules. (inset) The energy region up to 8~keV. The most prominent event clusters are the LEE and the two iron lines (purple dashed - K-$\alpha$, purple dotted - K-$\beta$). (main window) The energy region up to 0.5~keV, dominated by the LEE. The Li1's LEE is less prominent due to the cut based on light channel information, which removes the foil events. }
\label{fig:li2_spectrum}
\end{figure}

Our event selection for the blind data set of the Li1 module is visualised in figure~\ref{fig:li1_mainpars}. As the event selection was designed on the training data set, the remaining outliers are an effect of the imperfect generalisation from the training to the blind data. Nevertheless, overall we observe a good performance of our chosen cuts in the discrimination between recoils and artefacts.
In figure~\ref{fig:li2_spectrum} the calibrated spectrum of the final event selection can be seen.
At low energies the Li2 module has a significantly higher number of events compared to the Li1 module. This is related to a special class of events that is highlighted in figure~\ref{fig:li1_mainpars} (right, red). A significant share of the events with low recoil energies (below 1~keV) in the phonon channel coincidentally has the pulse shape of direct hits in the light channel (see section~\ref{sec_analysis_results}) and corresponds to large energy depositions. Direct hits feature a significantly sharper pulse shape as the phonon population is created instantaneously with a single particle scattering, while in the formation of a pulse shape from scintillation light multiple photons are collected and accumulate to form the observed pulse shape. In the past these events could be connected to the presence of the reflective foil inside the housing of the detector module \cite{bauer_thesis}. With the information from the light channel these events can be identified in the Li1 data, while in Li2, they remain in the final DM data set. 

\subsection{Efficiency and detector performance}
\label{subsec:efficiency}
\begin{figure}[!t]
\centering
\includegraphics[width=\linewidth]{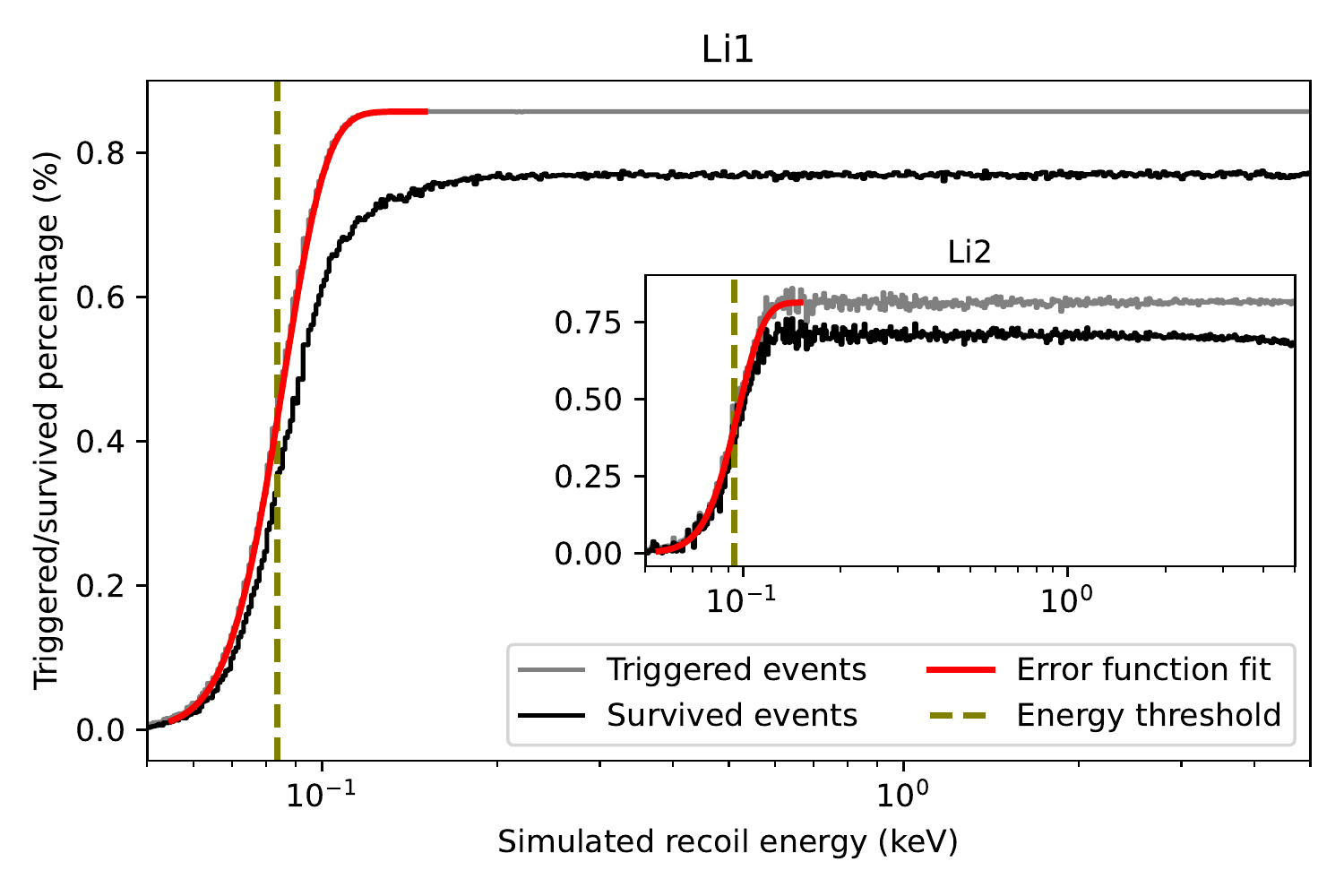}
\caption{The normalised trigger rate (grey) and survival rate (black) of simulated Li1 events (Li2 events in inset), as a function of the simulated recoil energy. The latter provides a realistic estimate of the survival probability. The energy threshold (olive, dashed) is the recoil energy at which the fitted error function (red) drops below 0.5 times the constant triggered fraction above threshold. The constant trigger efficiency for Li1 is (85.71 $\pm$ 0.01)\% and the trigger energy threshold  (83.60 $\pm$ 0.02)~eV. For Li2 the trigger efficiency is (81.26 $\pm$ 0.08)\% and the trigger energy threshold  (94.09 $\pm$ 0.13)~eV.}
\label{fig:li1_threshold}
\end{figure}

We evaluate the overall selection cut efficiency simulating 2$\times 10^6$ particle recoil events for the two modules respectively. These events are evenly distributed in the time of the measurement and the identical analysis chain of the blind data is applied to the simulated events. The energy dependent survival rate of the simulated events provides a realistic estimate of the survival probability of particle recoil events and is used for the limit calculation. To obtain a statement on the energy threshold, we fit an error function to the triggered events as a function of the simulated recoil energy. The energy threshold of our detectors is defined as the recoil energy at which the error function drops below half of its constant value at higher energies (see figure~\ref{fig:li1_threshold}). We find an energy threshold of (83.60~$\pm$~0.02)~eV for the Li1 module and (94.09~$\pm$~0.13)~eV for the Li2 module. These values correspond to the voltage value chosen as trigger threshold, converted to a recoil energy. The constant trigger efficiencies above threshold are (85.71~$\pm$~0.01)\% for Li1 and (81.26~$\pm$~0.08)\% for Li2. The plateau is not at unity due to the induced dead time from test and control pulses, and the dead time caused by previous triggers.

We estimate the baseline energy resolution of the phonon detector with the width of the fitted error function. This leads for the Li1 module to a value of (13.10~$\pm$~0.02)~eV, and for the Li2 module to a value of (15.89~$\pm$~0.18)~eV. For the light channel of the Li1 module, we estimate the baseline energy resolution by superimposing the standard event to a set of empty noise traces and measuring the standard deviation ($\sigma$) of the reconstructed pulse heights. We observe a baseline resolution of (748~$\pm$~7)~eV$_{ee}$ (ee energy scale) and (2.26~$\pm$~0.02)~eV (total energy scale). The method used for the phonon channel is more precise, as it includes also corrections of the detector response over time. However, the values agree with the ones obtained with the second method.

\begin{figure*}[!t]
\centering
\includegraphics[width=\textwidth]{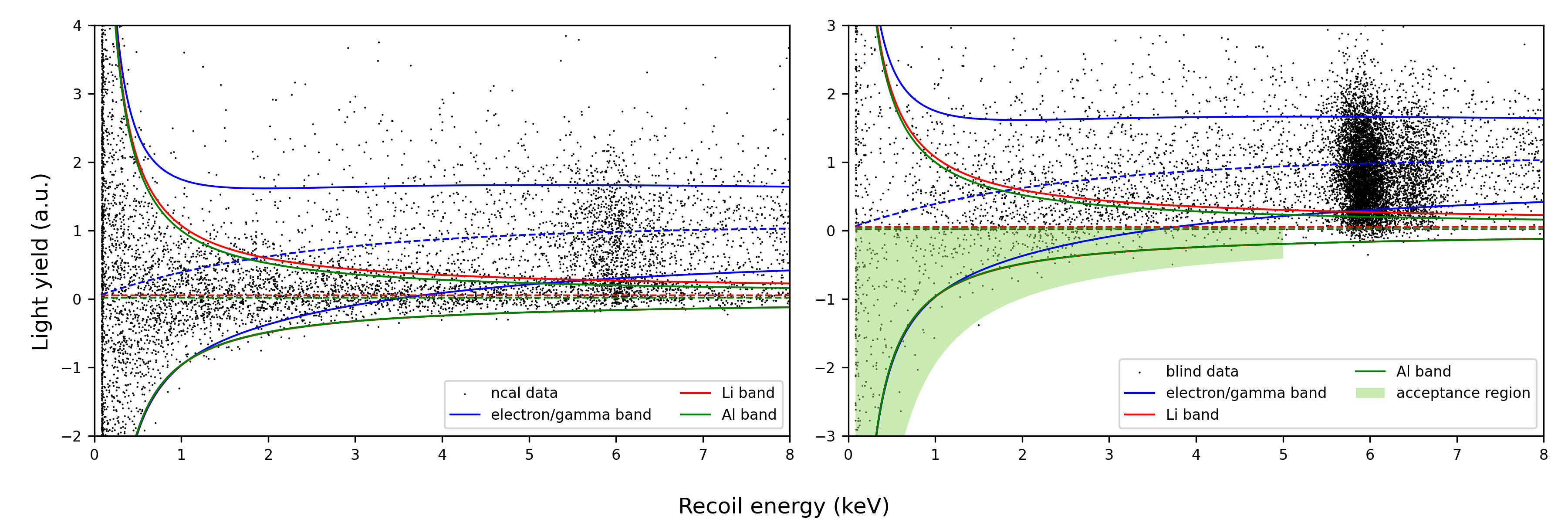}
\caption{Fitted light yield bands as a function of the recoil energy in the Li1 neutron (left) and blind (right) data sets, after application of the selection criteria discussed in section \ref{sec:eventselection}. Electron/$\gamma$ (blue) and nuclear recoils off the nuclei with odd proton number (lithium red, aluminium green) cluster in band-like structures and are fitted with Gauss distributions, with energy dependent means and standard deviations. The acceptance region for DM candidates (light green) is chosen as the lower half of the lithium and aluminium bands, mitigating the EM background.}
\label{fig:li1_bandplot}
\end{figure*}

\begin{figure*}[!]
\centering
\includegraphics[width=\textwidth]{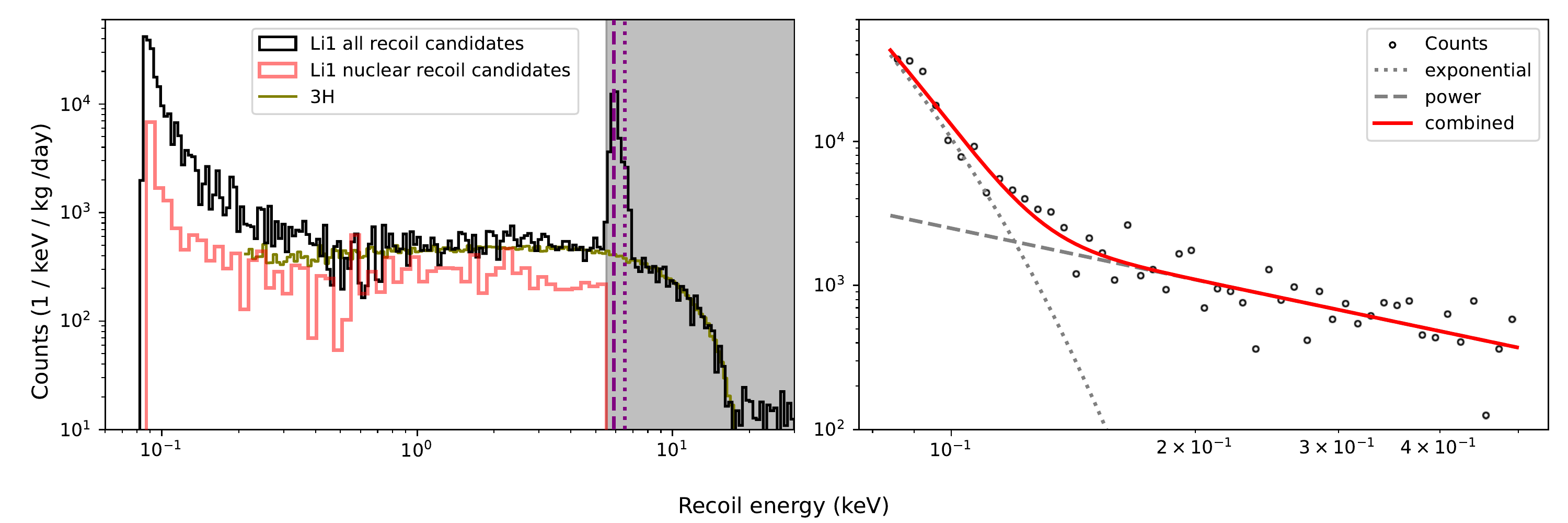}
\caption{Recoil energy spectrum of particle events inside the target of Li1. (left) The recoil spectrum up to 30~keV (black), the ROI for the DM search ends at 5.5~keV, indicated by the grey shade. The choice of the ROI is motivated in the text. The three prominent contributions are the clearly visible iron line (purple dashed - K-$\alpha$, purple dotted - K-$\beta$), the tritium background (olive line to guide the eye) and the LEE. The events within the acceptance region are considered nuclear recoil candidates (red). (right) The region below 0.5~keV, which is dominated by the LEE. The recoil energy spectrum of all recoil candidate events (black dots) can be fitted with the sum of an exponential (grey, dotted) and a power law component (grey, dashed). }
\label{fig:li1_spectrum}
\end{figure*}

\subsection{Results} \label{sec_analysis_results}


In the final DM data set, shown in figure~\ref{fig:li1_spectrum} (left, black), it is possible to identify three main contributions: first, the X-ray lines at 5.89 and 6.49~keV induced by the internal calibration source. Second, the beta spectrum of tritium. Beta events from tritium are expected due to the accumulation of tritium inside the crystal. $^6$Li has a high cross section for the reaction $^{6}Li(n,\alpha)$ which leaves behind tritium nuclei inside the crystal lattice. Third, a Low Energy Excess (LEE), a phenomenon that is seen by many experiments with low energy thresholds and is the matter of ongoing discussion in the community \cite{https://doi.org/10.48550/arxiv.2202.05097, PhysRevD.102.015017, https://doi.org/10.48550/arxiv.2112.14495, https://doi.org/10.48550/arxiv.2202.03436}. Its origin is still unclear. An interpretation of the foil events as the origin of the LEE can be excluded due to their significantly different spectral shape. As there is no method to discriminate particle recoils from LEE events, we treat them as particle recoils in the analysis. 

In figure~\ref{fig:li1_spectrum} (right) the count rate between threshold and 500~eV is displayed in this energy region the main contribution is given by the LEE. The LEE spectrum can be fitted with a combination of an exponential and a power law function:
\begin{equation}\label{eq:fit}
f(x,a,b,c,d) = a \exp(-bx) + c x^{-d},
\end{equation}
\noindent where x is the running parameter, and a, b, c and d are free fit parameters. 
The values obtained with a $\chi^2$ fit to the binned spectrum for the Li1 LEE are summarised in table~\ref{tab:li1_spectra_fit} to make a comparison with spectra obtained from other measurements possible.


The region of interest for a DM analysis is defined using the light yield parameter (LY)

\begin{align}
    LY = \frac{E_l}{E_p},
    \label{eq:ly}
\end{align}

\noindent which quantifies the collected scintillation light from an individual event. Here, $E_l$ is the energy of the light channel in ee-energy units, and $E_p$ the energy of the phonon channel in total energy units. Note that our definition of the LY automatically normalises it to one for recoils induced by the iron source. The amount of produced scintillation photons in the target is quenched for nuclear recoils, with respect to electron and gamma recoils (EM recoils). We use this information to suppress the EM background. The quenching factor is measured in-situ with the neutron calibration data.
The exposure of the neutron calibration data set is 0.178~kg~days. The same analysis chain for the blind data set has been applied to the neutron calibration data except that no coincidence cuts were applied, to keep higher statistics.

Figure~\ref{fig:li1_bandplot} shows the LY versus energy from neutron calibration and blind data. The nuclear recoils are quenched according to the mass of the nucleus which they scatter off. For the two lithium isotopes only one band is drawn since no big difference is expected due to the negligible difference in masses. The oxygen band is not drawn, because it overlaps almost fully with the aluminium band. We can clearly identify the band of neutrons which scatter on nuclei, while the EM band is wider and less prominently pronounced. However, the light yield of the iron source, which builds a clearly visible cluster around 6~keV, indicates the position of the EM bands. To quantify the position of the nuclear recoil bands an unbinned likelihood fit of the recoil bands is performed: each band is described by a Gaussian distribution with energy dependent mean and standard deviation. Their parametrisation is non-trivial and described in detail in reference~\cite{Schmiedmayer:thesis}. The bands plotted in figure~\ref{fig:li1_bandplot} correspond to the 80\% central interval of the Gauss function. The lower half of the lithium nuclear recoil band is defined as the region of interest (ROI, acceptance region) for the DM search, as a trade-off between efficiency and background minimisation. Our ROI ends below the iron line i.e.~it extends from the trigger threshold of 83.60~eV (Li1) and 94.09~eV (Li2) to 5.5~keV. We do not consider higher recoil energies for our DM search for two reasons: first, the region does not show a significant signal expectation for recoils from sub-GeV/c$^2$ DM particles, and second, it is dominated by the contribution from the iron source. 
The resulting energy spectrum of events that fall in the ROI is shown in figure~\ref{fig:li1_spectrum} (left, red).


\begin{table}[!t]
\centering
\begin{tabular}{@{}llll@{}} \toprule
 & value & uncertainty & units \\
 \hline
 $\textbf{a}$ &  $4.7 \cdot 10^8$  & $ \pm$ 7.3$  \cdot 10^8$ & $(\text{keV} \cdot \text{kg} \cdot \text{day})^{-1}$\\
$\textbf{b}$ & $84$ & $\pm$ 16 & $(\text{keV})^{-1}$\\
$\textbf{c}$   &  $162$ & $\pm$ 41 & $(\text{keV}^{\text{(1 - d)}} \cdot \text{kg} \cdot \text{day})^{-1}$ \\
$\textbf{d}$ &  $1.2$ & $\pm$ 0.2 &\\
\end{tabular}

\caption{The parameters obtained from a $\chi^2$ fit of Eq. (\ref{eq:fit}) to the binned spectrum for the Li1 LEE.}
\label{tab:li1_spectra_fit}
\end{table}

\section{Dark matter results}
\label{sec:Results}

\begin{figure*}[!t]
\centering
\includegraphics[width=\linewidth]{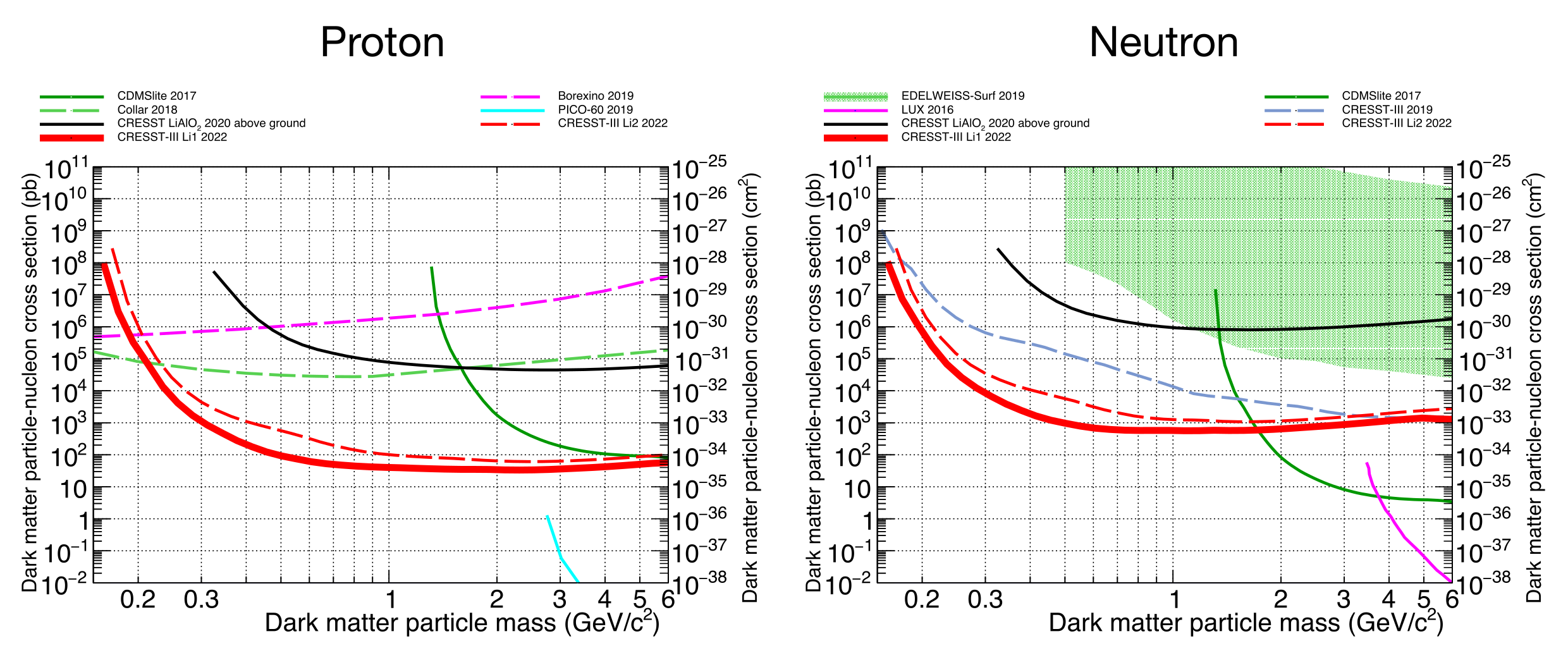}
\caption{The exclusion limits for \textit{proton-only} (left) and \textit{neutron-only}(right) spin-dependent DM-nucleus cross sections versus DM particle mass set by various experiments compared with the two lithium modules described in this work with \textsuperscript{6}Li, \textsuperscript{7}Li and \textsuperscript{27}Al. This work gives the most stringent limits between 0.25~GeV/c\textsuperscript{2} - 2.5~GeV/c\textsuperscript{2} for \textit{proton-only} and between 0.16~GeV/c\textsuperscript{2} - 1.5~GeV/c\textsuperscript{2} for \textit{neutron-only} interactions. The solid red line shows the Li1 limits which includes the scintillation light information and the dashed red line shows the Li2 limits where no light information was available (hence worse). The previous above ground results from CRESST using the same detector material and procedure with higher threshold and lower exposure is also shown with solid black line\cite{angloher_probing_2022}. Also, CRESST-III 2019 results for \textit{neutron-only} interactions  using \textsuperscript{17}O are shown also in dashed light-blue line(right)\cite{PhysRevD.100.102002}. Additionally, we show the limits from other experiments: EDELWEISS \cite{Edelweiss} and CDMSlite with \textsuperscript{73}Ge\cite{CDMSlite}, PICO with \textsuperscript{19}F\cite{Pico}, LUX \cite{LUX} which use \textsuperscript{129}Xe + \textsuperscript{131}Xe, J. I. Collar with \textsuperscript{1}H \cite{Collar_2018} and the constrain derived in \cite{Bringmann_2019} from Borexino.} 
\label{fig:limits}
\end{figure*}

In order to calculate the upper limit for spin-dependent DM-nucleon interactions, we work in the limit of zero momentum transfer and thus neglect the form factors. The expected different recoil rate for the \textit{proton/neutron only} spin-dependent DM interactions is given by

\begin{align}
\begin{split}
    \frac{dR}{dE_R} = \frac{2\rho_0}{m_{\chi}} \sigma_{p/n}^{SD} \sum_{i,T} &f_{i,T} \left(\frac{J_{i,T} + 1}{3J_{i,T}}\right)\cdot\\ &\left(\frac{\braket{S_{p/n,i,T}}^2}{\mu^2_{p/n}}\right) \eta(v_{min}),
    \label{eq:diff_rate}
\end{split}
\end{align}
\noindent where $E_R$ is the recoil energy; $\rho_0$ is the local DM density and $m_\chi$ is the WIMP mass and $\sigma_{p/n}^{SD}$ is the reference DM-proton/neutron cross section. The parameter $f_{i,T}$ is the fraction of each nucleus in the target scaled by its mass and is given by 

\begin{equation}
    f_{i,T} = \frac{n_T \zeta^i m^i_T}{\sum_{i,T'} n_{T'} \zeta^i m^i_{T'}},
\end{equation}

\noindent where $n_T$ is the multiplicity of nucleus $T$, $\zeta^i$ is the natural abundance of isotope $i$, and $m^i_T$ is its mass. It should be noted that we consider $f_{oxygen}=0$, i.e.~we do not include the contribution from oxygen in the spin-dependent interaction. The reason being very low natural abundance of $^{17}$O (i.e. 0.0367\%) and thus including it changes the expected DM rate only negligibly. Furthermore, $J_{i,T}$ is the nuclear ground state angular momentum of the isotope $i$ of nucleus $T$; $\braket{S_{p/n,i,T}}$ is the expected value of the \textit{proton/neutron} spins in the target isotope $i$ of nucleus $T$ and $\mu^2_{p/n}$ the nucleon-DM reduced mass; and $\eta(v_{min})$ is the mean inverse velocity in the standard model halo \cite{RevModPhys.85.1561} where $v_{min}$ is the minimum velocity required to produce a nuclear recoil of energy $E_R$ \cite{Lee_2013}. This formalism is equivalent to the one that was employed in our previous work~\cite{abdelhameed_first_2019}. 

We adopt the standard DM halo model that assumes a Maxwellian velocity distribution and a local DM density of $\rho_{\textup{DM}}$ = 0.3~(GeV/c\textsuperscript{2})/cm\textsuperscript{3} \cite{refId0}, the galactic escape velocity at the position of the sun of $\upsilon_{\textup{esc}}$ = 544~km/s \cite{10.1111/j.1365-2966.2007.11964.x}, and the solar orbital velocity of $\upsilon_\odot$ = 220~km/s \cite{10.1093/mnras/221.4.1023}. For the calculation of \textit{neutron-only} and \textit{proton-only} limits, we use $\braket{S_\textup{n}}$ = $\braket{S_\textup{p}}$ = 0.472  for \textsuperscript{6}Li \cite{PhysRevC.102.014001}, $\braket{S_\textup{p}}$ = 0.497 for \textsuperscript{7}Li \cite{Pacheco1989} and $\braket{S_\textup{n}}$ = 0.0296, $\braket{S_\textup{p}}$ = 0.343 for \textsuperscript{27}Al \cite{Engel1995}. 

For the calculation of DM exclusion limits it needs to be understood how a DM signal would look like after application of our analysis chain. For this, the input simulated spectrum discussed in section~\ref{subsec:efficiency} is re-weighed such that it resembles the expected recoil spectrum from each DM mass. The resulting recoil spectrum seen after triggering, applying data quality cuts, and energy reconstruction in the same way it is done for the blind set, automatically includes the information about the detector resolution and threshold. Thus, an \textit{observed} energy spectrum for a given \textit{injected} spectrum is obtained. We additionally remove events where the reconstructed amplitudes differs from the injected amplitude by more than three times the detector resolution. This is done in order to avoid any non-physical reconstruction of sub-threshold events if they happen to pile-up with exceptionally strong upward fluctuations of the noise baseline. The same formalism was also employed and discussed in our previous work \cite{PhysRevD.100.102002}.


The choice of ROI is motivated in section~\ref{sec_analysis_results}, where we define our candidate events. A similar procedure was used for the calculation of the limits from the Li2 blind data except that no band fit could be employed. The exclusion limits are finally calculated using Yellin's optimum interval method \cite{Yellin_2002,Yellin_2007} to extract the upper limit on the cross section of DM particles with \textsuperscript{6}Li along with \textsuperscript{7}Li and \textsuperscript{27}Al. Limits on the spin-dependent reference cross section for \textit{proton/neutron-only} interaction are shown in figure~\ref{fig:limits} for DM masses from 0.16 to 6~GeV/c\textsuperscript{2}, for both the modules, and compared with those from other experiments. These results are reported using Yellin's optimum interval method to extract the 90\% confidence level upper limits. We can see around 3-4 orders of magnitude improvement in both proton and neutron limits for the entire probed mass range, compared to our previous test done with the same material in the above ground facility with a higher energy threshold and lower exposure \cite{angloher_probing_2022}. The Li1 module provides up to an order of magnitude better results than the Li2 module because of the additional scintillation light information. For very low masses, which are dominated by the LEE that cannot be discriminated from nuclear recoils, the difference is negligible. For the \textit{proton-only} interactions, we improve the existing limits from 0.25 to 2.5~GeV/c\textsuperscript{2} by up to a factor of 2.5 compared to other experiments. For the \textit{neutron-only} interactions, we achieve the strongest limit between 0.16 and 1.5~GeV/c\textsuperscript{2}, more than an order of magnitude better than the limits from our 2019 results using \textsuperscript{17}O \cite{PhysRevD.100.102002}. 

\section{Conclusion}
\label{sec:Conclusion}

In this work we present the detailed analysis and results of two lithium-based cryogenic detectors operated in the underground facility of the CRESST experiment at LNGS. We highlight the results of the best performing one and validate its analysis with the result of a second, identically manufactured detector. The best performing one achieves a threshold of 83.60~eV that corresponds to sensitivities down to a DM mass of 0.16~GeV/c\textsuperscript{2}. We have probed spin-dependent DM particle interactions with nuclei, distinguishing \textit{proton-only} and \textit{neutron-only} interactions. For \textit{proton-only} interactions, leading exclusion limits for the mass region between 0.25 and 2.5~GeV/c$^2$ are presented. Additionally, for \textit{neutron-only} interactions, best sensitivity was achieved in the mass range of 0.16 and 1.5~GeV/c$^2$.

The results of this run showed that LiAlO$_2$ is an excellent material to study spin-dependent interactions and will therefore be included in future CRESST projects. Below DM masses of 0.6~GeV/c$^2$ the limit-setting power of the CRESST lithium detector modules decreases. The reason for this is an excess of events at low energies. The source of these is currently under investigation. 

\section*{Acknowledgements}
We are grateful to LNGS for their generous support of CRESST. This work has been funded by the Deutsche Forschungsgemeinschaft (DFG, German Research Foundation) under Germany's Excellence Strategy – EXC 2094 – 390783311 and through the Sonderforschungsbereich (Collaborative Research Center) SFB1258 ‘Neutrinos and Dark Matter in Astro- and Particle Physics’, by the BMBF 05A20WO1 and 05A20VTA and by the Austrian Science Fund (FWF): I5420-N, W1252-N27 and FG1 and by the Austrian research promotion agency (FFG), project ML4CPD. The Bratislava group acknowledges a partial support provided by the Slovak Research and Development Agency (project APVV-15-0576). The computational results presented were partially obtained using the Vienna CLIP cluster and the Munich MPCDF.

\bibliographystyle{h-physrev}
\bibliography{main}
\end{document}